\documentclass[a4paper,twoside]{article}

\usepackage{epsfig}
\usepackage{subcaption}
\usepackage{calc}
\usepackage{amssymb}
\usepackage{amstext}
\usepackage{amsmath}
\usepackage{amsthm}
\usepackage{multicol}
\usepackage{pslatex}
\usepackage{apalike}

\usepackage[hidelinks]{hyperref}

\usepackage{SCITEPRESS}     

\theoremstyle{definition}
\newtheorem*{definition*}{Definition}
\newtheorem*{algorithm*}{Algorithm}
\newtheorem{remark}{Remark}
\theoremstyle{plain}
\newtheorem*{theorem*}{Theorem}
\newtheorem*{corollary*}{Corollary}

\begin{document}

\title{Characteristics-based Simulink implementation of first-order quasilinear partial differential equations}

\author{\authorname{Anton Ponomarev\sup{1}\orcidAuthor{0000-0002-8753-9759}, Julian Hofmann\sup{2}\orcidAuthor{0000-0003-3210-4368} and Lutz Gr{\"o}ll\sup{2}\orcidAuthor{0000-0003-3433-1425}}
\affiliation{\sup{1}Saint Petersburg State University, 7/9 Universitetskaya nab., St. Petersburg, 199034 Russia}
\affiliation{\sup{2}Karlsruhe Institute of Technology, Institute for Automation and Applied Informatics,\\ Hermann-von-Helmholtz-Platz 1, 76344 Eggenstein-Leopoldshafen, Germany}
\email{anton.pon.math@gmail.com, \{julian.hofmann, lutz.groell\}@kit.edu}}


\abstract{The paper deals with solving first-order quasilinear partial differential equations in an online simulation environment, such as Simulink, utilizing the well-known and well-recommended method of characteristics. Compared to the commonly applied space discretization methods on static grids, the characteristics-based approach provides better numerical stability. Simulink subsystem implementing the method of characteristics is developed. It employs Simulink's built-in solver and its zero-crossing detection algorithm to perform simultaneous integration of a pool of characteristics as well as to create new characteristics dynamically and discard the old ones. Numerical accuracy of the solution thus obtained is established. The subsystem has been tested on a full-state feedback example and produced better results than the space discretization-based ``method of lines''. The implementation is available for download and can be used in a wide range of models.}

\onecolumn \maketitle \normalsize \setcounter{footnote}{0} \vfill

\section{\uppercase{Introduction}}
\label{sec:Intro}

\noindent Dynamical systems described by first-order quasilinear partial differential equations (PDEs) are commonly interpreted as convective-reactive processes, i.e., they represent transport phenomena with sources and sinks. Such phenomena play an important role in chemical and process engineering, traffic flow models, material science, biology, etc. \cite{mahmoudi2019convective,kachroo2018traffic,ingham1998transport,truskey2004transport}. PDEs of this form typically arise as mathematical formulation of the physical conservation laws of spatially distributed quantities (mass, momentum, etc.) applied locally (to an infinitesimal volume) in which case they are also known as continuity equations \cite{greenkorn2018momentum}. 

In this paper we discuss computer simulation of arbitrary convection-reaction dynamics described by the following first-order quasilinear PDE on a bounded spatial domain $[0, \ell]$:
\begin{subequations} \label{eq: general case}
    \begin{gather}
        w_t + v \big( t, x, w(t, \cdot)_{[0, \ell]} \big) w_x =
            f \big( t, x, w(t, \cdot)_{[0, \ell]} \big),
            \label{eq: general case pde} \\
        w(0, x) = w_0(x), \; x \in [0, \ell], \\
        w(t, 0) = u(t)
    \end{gather}
\end{subequations}
where $t \geq 0$, $x \geq 0$, $\ell \in (0, \infty)$, $w(t, x)$ is the internal state of the system, $w_t$ and $w_x$ are partial derivatives of $w$ at the point $(t, x)$, $w_0$ is an $L^2$ (square-integrable) initial state function, $v$ represents the convection velocity ($v \geq \varepsilon > 0$ for all possible arguments), $f$ can be interpreted as a source, sink, or reaction term, and $u$ plays the role of a boundary input into the system. Symbol $w(t, \cdot)_{[0, \ell]}$ stands for the restriction of the function $w(t, x)$ to $x \in [0, \ell]$.

The terms $v$ and $f$ in (\ref{eq: general case}) depend on time $t$, spatial position $x$, and full system state $w(t, \cdot)_{[0, \ell]}$. In addition, they may also depend on other signals, e.g., on the output of a dynamic feedback controller. The same holds true for the boundary input $u(t)$.

\begin{remark} \label{re: domain}
	From the theoretical standpoint, it would be enough to define functions $v$ and $f$ only for $x \in [0, \ell]$. However, our simulation algorithm described below requires that the system be defined on a larger domain $x \in [0, \bar\ell]$ with some $\bar\ell > \ell$ (see Remark \ref{re: domain explanation}). Although a sufficiently large but finite $\bar\ell$ could be estimated, for the sake of brevity we opt to assume that the system is defined for all $x \geq 0$. In the practical sense, it means that the functions $v$ and $f$ should be extended from $x \in [0, \ell]$ to $x > \ell$ in a physically meaningful way.
\end{remark}

\begin{remark}
    The system (\ref{eq: general case}) admits unique classical solution (i.e., well-defined single-valued function) if there are no intersecting characteristics \cite{polyanin2002handbook}. Otherwise, the so-called shock waves appear which can be dealt with using the concept of multi-valued solutions or by introducing discontinuity at the shock wave's front. In this paper we avoid making restrictive a priori assumptions that would exclude the possibility of intersecting characteristics. Furthermore, our algorithm will not handle the shock waves by itself. Instead, the algorithm will detect intersecting characteristics during runtime and inform the user of the problem.
\end{remark}

Let us discuss the existing approaches to computer simulation of the dynamical system (\ref{eq: general case}) in the MATLAB/Simulink environment paying primary attention to the method of characteristics \cite{polyanin2002handbook} as it is arguably the most accurate and efficient tool to solve a general PDE like (\ref{eq: general case pde}).

There are numerous MATLAB scripts that implement the method of characteristics, e.g., \cite{quasilinm}. Assuming that the functions $v$, $f$, and $u$ are given as functions of $t$, $x$, and (to some extent) $w(t, \cdot)_{[0, \ell]}$, such MATLAB scripts can be used to calculate solutions of the systems of the form (\ref{eq: general case}). However, these solvers are ``offline'' in the sense that the state $w(t, x)$ is available for all values of $t$ and $x$ only after the script is executed completely. We cannot, in a straightforward manner, embed such an ``offline'' solver into a Simulink model as a \textit{MATLAB Function} block because it will not output the full state $w(t, \cdot)_{[0, \ell]}$ as the simulation time $t$ increases.

We are interested in an ``online'' solver, which is to say that, as the simulation time $t$ increases, it should continuously produce an approximation of the current state $w(t, \cdot)_{[0, \ell]}$. An ``online'' solver is required if at each moment in time the terms $v$, $f$, and $u$ in (\ref{eq: general case})  depend on the full system state $w(t, \cdot)_{[0, \ell]}$ in a non-trivial manner or are defined via feedback of the full system state through another dynamical system which may be the case, e.g., under feedback control.

Simulink is a suitable environment to implement (\ref{eq: general case}) in the ``online'' fashion. Having a block which takes the values of $v$, $f$, and $u$ and outputs a discrete approximation of the full system state $w(t, \cdot)_{[0, \ell]}$, it would be easy to plug it into a larger system. However, we are not aware of Simulink implementations of the method of characteristics for the general case of system (\ref{eq: general case}). Let us describe some alternatives that are available in Simulink at the moment and motivate our effort to implement the method of characteristics.

Space discretization (finite difference) method, also known as the \emph{Method of Lines} \cite{schiesser2012numerical}, can be used to transform the PDE into a system of ordinary differential equations (ODEs) which has native Simulink implementation. In general, these approaches suffer from instability because of numerical diffusion and dispersion \cite{Zijlema15}. We illustrate this problem in Section \ref{sec: example}.

Another approach is to reformulate the PDE as a time-delay system. For example, \cite{witrant2010modeling} consider a variable-velocity transport system with sinks or sources
\begin{equation} \label{eq: Witrant case}
    \begin{gathered}
        w_t + v(t) w_x = f(t), \\
        w(t, 0) = u(t), \\
        y(t) = w(t, 1).
    \end{gathered}
\end{equation}
Using the method of characteristics, it is transformed into a delay-differential equation which can be plugged into a Simulink model. The special case of (\ref{eq: Witrant case}) with $f(t)=0$ is included in Simulink under the name of \emph{Variable Transport Delay} block \cite{ZhangYeddanapudi2012}. In general, however, converting a model like (\ref{eq: general case}) into a time-delay form requires deep analysis and may not always be possible.

Finally, there are examples of connecting Simulink to dedicated PDE solving environments \cite{van2009integrated} but such an approach is only justified for simulating complex phenomena. For a simple transport process the overhead is excessive.

Hence, we set the goal of designing a simple but versatile Simulink block which can simulate the dynamics of a wide range of first-order quasilinear PDEs based on the method of characteristics. The block is to be employed in modeling the systems like (\ref{eq: general case}). The accuracy of simulation should be adjustable via some parameters. The block has to be easy to use without extensive analysis of the mathematical model. Furthermore, it must have straightforward interface and be easily embeddable into larger Simulink models.

\begin{remark}
    To the best of the authors' knowledge, the only example that comes close to reaching the goal of this paper is \cite{Herran-GonzalezCruzAndres-ToroEtAl2009}. It is a library of Simulink blocks modeling gas ducts, valves, compressors, etc. Although the blocks indeed implement the method of characteristics, the library is directed exclusively at modeling gas distribution pipeline networks which results in a limited choice of possible dynamics. Our approach is more general as we model dynamics (\ref{eq: general case}) in an abstract sense. We allow arbitrary velocity and source terms as well as initial functions.
\end{remark}

The plan of the paper is as follows. In Sec.~\ref{sec: method} we provide theoretical background to the proposed simulation approach. Specifically, in Sec.~\ref{subsec: characteristics} we recall the method of characteristics whereas in Sec.~\ref{subsec: simulation} we introduce its algorithmic realization suitable for ``online'' simulation and assess the accuracy of the algorithm. Sec.~\ref{sec: implementation} describes implementation of the algorithm as a Simulink block. In Sec.~\ref{sec: example}, performance of the block is compared with the method of lines using an illustrative example.

\section{\uppercase{Method}}
\label{sec: method}

\subsection{The Method of Characteristics}
\label{subsec: characteristics}

\noindent Our approach to simulation of the dynamics (\ref{eq: general case}) is based on the method of characteristics \cite{polyanin2002handbook}. Let us recall the definition of the characteristics.

\begin{definition*}
    The \emph{characteristics} of the system (\ref{eq: general case}) starting at the point $(t_0, x_0)$ on the initial boundary ($t_0 = 0$ and $x_0 \geq 0$) or on the input boundary ($t_0 > 0$ and $x_0 = 0$) are the solutions $\xi(t; t_0, x_0)$ and $\omega(t; t_0, x_0)$ of the system of ordinary differential equations (ODEs)
    \begin{equation} \label{eq: characteristics}
        \begin{aligned}
            \dot\xi(t; t_0, x_0) &=
                v \big( t, \xi(t; t_0, x_0),
                w(t, \cdot)_{[0, \ell]} \big), \\
            \dot\omega(t; t_0, x_0) &=
                f \big( t, \xi(t; t_0, x_0),
                w(t, \cdot)_{[0, \ell]} \big)
        \end{aligned}
    \end{equation}
    where $t \geq 0$, with initial conditions
    \begin{equation*}
        \xi(0; 0, x_0) = x_0, \quad
        \omega(0; 0, x_0) = w_0(x_0), \quad
        x_0 \geq 0
    \end{equation*}
    (in the case of initial boundary) or
    \begin{equation*}
        \xi(t_0; t_0, 0) = 0, \quad
        \omega(t_0; t_0, 0) = u(t_0), \quad
        t_0 > 0
    \end{equation*}
    (in the case of input boundary). Here $\dot\xi$ and $\dot\omega$ denote the derivatives of $\xi$ and $\omega$ with respect to $t$.
\end{definition*}

The functions $\xi$ and $\omega$ being a solution of (\ref{eq: characteristics}), we know from \cite{polyanin2002handbook} that the solution $w(t, x)$ of (\ref{eq: general case}) takes the values $\omega(t; t_0, x_0)$ along the curve $\big( t, \xi(t; t_0, x_0) \big)$, i.e.,
\begin{equation} \label{eq: solution from characteristic}
    w \big( t, \xi(t; t_0, x_0) \big) =
        \omega(t; t_0, x_0).
\end{equation}

Thus, starting at a number of points $(t_0, x_0)$ on the initial and input boundaries and integrating the characteristic ODEs (\ref{eq: characteristics}) one can find the solution of the PDE problem (\ref{eq: general case}) along a number of curves spanning the domain. However, the structure of (\ref{eq: general case}) is such that the full state $w(t, \cdot)_{[0, \ell]}$ of the PDE is involved in the characteristic ODEs (\ref{eq: characteristics}). Therefore, the ODEs (\ref{eq: characteristics}) pertaining to different characteristics are to be integrated together (in parallel, so to speak) while the function $w(t, \cdot)_{[0, \ell]}$ is interpolated using the relations (\ref{eq: solution from characteristic}). In the next Section we clarify this approach.

\subsection{Simulation Algorithm}
\label{subsec: simulation}

\noindent Our proposal is to start simulation at $t = 0$ with a set of characteristics spread across the spatial domain and integrate their respective ODEs (\ref{eq: characteristics}) in parallel. To this end, the full state $w(t, \cdot)_{[0, \ell]}$ must be plugged into the right-hand side of (\ref{eq: characteristics}). However, the full state is unknown. Only its values on the characteristic curves are available, according to (\ref{eq: solution from characteristic}). Therefore, we will approximate $w(t, \cdot)_{[0, \ell]}$ in (\ref{eq: characteristics}) via interpolation between the characteristics. Once a characteristic has moved far enough out of the spatial domain and has no effect on the full PDE state anymore, it is removed from the set. New characteristics are created on the input boundary and added to the set when appropriate according to some rules given below.

To give a precise description of the algorithm, let us represent the set of characteristics as two variable-length vectors $\Xi(t)$ and $\Omega(t)$ each of which at time $t$ contains $N(t) \geq 2$ elements:
\begin{equation} \label{eq: Xi and Omega}
    \begin{aligned}
        \Xi(t) &=
            \big( \xi_1(t), \xi_2(t), \dots, \xi_{N(t)}(t) \big), \\
        \Omega(t) &=
            \big( \omega_1(t), \omega_2(t), \dots, \omega_{N(t)}(t) \big).
    \end{aligned}
\end{equation}
These vectors are to be handled by the following hybrid algorithm which combines continuous integration with event-triggered state resets.

\begin{algorithm*} Given parameters $\Delta_x \in (0, \ell)$, $\Delta_w > 0$, and $\Delta_t > 0$, the rules for the simulation of (\ref{eq: general case}) are:
    \begin{enumerate}
        \item \emph{Initialization}: initial number of characteristics $N(0) \geq 2$ and the elements of the vectors $\Xi(0)$ and $\Omega(0)$ are selected to approximate the initial function such that
        \begin{equation} \label{eq: characteristics init}
            \begin{gathered}
                0 = \xi_1(0) < \xi_2(0) < \dots < \xi_{N(0)}(0) = \ell, \\
                \omega_i(0) = w_0 \big( \xi_i(0) \big)
                    \quad \big( i = 1, 2, \dots, N(0) \big)
            \end{gathered}
        \end{equation}
        and
        \begin{equation*}
            \begin{aligned}
                \xi_{i+1}(0) - \xi_i(0) &\leq \Delta_x, \\
                |w_0(x) - \omega_i(0)| &\leq \Delta_w \\
                \big( x \in [\xi_i(0), \xi_{i+1}(0)], \quad
                    i = &1, 2, \dots, N(0)-1 \big).
            \end{aligned}
        \end{equation*}
        Moreover, an auxiliary variable $t_\text{LC}(t)$, denoting the last time when new characteristic was created, is initialized as $t_\text{LC}(0) = 0$.
        \item \emph{Dynamics} (continuous integration): $\Xi(t)$ and $\Omega(t)$ evolve according to the equations similar to (\ref{eq: characteristics}):
        \begin{equation} \label{eq: characteristics algo}
            \begin{aligned}
                \dot\xi_i(t) &=
                    v \big( t, \xi_i(t),
                    \tilde w(t, \cdot)_{[0, \ell]} \big), \\
                \dot\omega_i(t) &=
                    f \big(t, \xi_i(t),
                    \tilde w(t, \cdot)_{[0, \ell]} \big) \\
                &\big( i = 1, 2, \dots, N(t) \big)
            \end{aligned}
        \end{equation}
        with initial conditions (\ref{eq: characteristics init}). The ODEs (\ref{eq: characteristics algo}) are obtained from (\ref{eq: characteristics}) by substitution, in place of the unknown $w(t, \cdot)_{[0, \ell]}$, its approximation $\tilde w(t, \cdot)_{[0, \ell]}$. The latter function is an interpolant over the grid consisting of a point on the input boundary and $N(t)$ points on the characteristic curves, as suggested by (\ref{eq: solution from characteristic}):
        \begin{equation} \label{eq: grid}
            \begin{aligned}
                \tilde w(t, 0) &= u(t), \\
                \tilde w \big(t, \xi_i(t) \big) &= \omega_i(t) \quad
                    \big( i = 1, 2, \dots, N(t) \big).
            \end{aligned}
        \end{equation}
        Depending on the expected analytic properties of the solution, a suitable interpolation scheme should be chosen here: nearest neighbor, linear, spline, etc.
        
        The values $N(t)$ and $t_\text{LC}(t)$ are kept constant during integration of the ODEs.
        \item \emph{Removal trigger} (state reset): fulfillment of the condition
        \begin{equation} \label{eq: trigger remove}
            \xi_{N(t)-1}(t) \geq \ell
        \end{equation}
        triggers removal of the oldest ($N(t)$'th) characteristic from the set:
        \begin{equation} \label{eq: procedure remove}
            N(t+0) := N(t) - 1
        \end{equation}
        where the argument $(t+0)$ indicates the updated value of the variable during the state reset.
        \item \emph{Creation triggers} (state reset): fulfillment of any of the conditions
        \begin{subequations} \label{eq: trigger create}
            \begin{align}
                \xi_1(t) &\geq \Delta_x, \label{eq: trigger create x} \\
                |\omega_1(t) - u(t)| &\geq \Delta_w, \\
                t - t_\text{LC}(t) &\geq \Delta_t
            \end{align}
        \end{subequations}
        triggers creation of a new characteristic at the input boundary and resetting $t_\mathrm{LC}$:
        \begin{equation*} 
            \begin{aligned}
                N(t+0) &:= N(t) + 1, \\
                \xi_1(t+0) &:= 0, \\
                \xi_{i+1}(t+0) &:= \xi_i(t), \\
                \omega_1(t+0) &:= u(t), \\
                \omega_{i+1}(t+0) &:= \omega_i(t), \\
                t_\text{LC}(t+0) &:= t \\
                \big( i = 1, 2, &\dots, N(t) \big).
            \end{aligned}
        \end{equation*}
        \item \textit{Output}: the algorithm outputs vectors $\Xi(t)$ and $\Omega(t)$ which can be used to construct an approximation $\tilde w(t, \cdot)_{[0, \ell]}$ of the full PDE state $w(t, \cdot)_{[0, \ell]}$ by interpolation over the grid (\ref{eq: grid}).
    \end{enumerate}
\end{algorithm*}

\begin{remark}
    Assuming exact trigger detection, the algorithm preserves the condition $N(t) \geq 2$ because characteristic creation trigger (\ref{eq: trigger create x}), thanks to the requirement $\Delta_x < \ell$, ensures that $\xi_1(t) < \ell$ for all $t$ which guarantees that removal condition (\ref{eq: trigger remove}) cannot be satisfied if $N(t) = 2$. Thus, $N(t)$ cannot drop below 2.
\end{remark}

Straightforward estimations of the solutions of the characteristic equations (\ref{eq: characteristics}) yield the following statements regarding the accuracy of the Algorithm.

\begin{theorem*}
    Suppose the PDE (\ref{eq: general case pde}) has the form
    \begin{equation}\label{eq: Theo pde}
        w_t + v \big( t, x, w(t, x) \big) w_x
            = f \big( t, x, w(t, x) \big)
    \end{equation}
    and there exist positive constants $\hat T$ and $\hat F$ such that the inequalities
    \begin{equation*}
        \begin{gathered}
            v(t, x, w) \geq \frac{\ell}{\hat T}, \\
            \begin{Vmatrix}
                v(t, \tilde x, \tilde w) - v(t, x, w) \\
                f(t, \tilde x, \tilde w) - f(t, x, w)
            \end{Vmatrix} \leq \hat F
            \begin{Vmatrix}
                \tilde x - x \\
                \tilde w - w
            \end{Vmatrix}
        \end{gathered}
    \end{equation*}
    hold for all values of $t, x, \tilde x \geq 0$ and $w, \tilde w$ where $\Vert\cdot\Vert$ denotes Euclidean vector norm in $\mathbb{R}^2$. Then, assuming that ODEs (\ref{eq: characteristics algo}) are integrated exactly and triggers (\ref{eq: trigger remove}), (\ref{eq: trigger create}) are detected perfectly on time, the following estimation is valid for all $t\geq 0$ and $x \in [\xi_i(t), \xi_{i+1}(t)]$, $i = 1, 2, \dots, N(t)-1$:
    \begin{equation*} 
        \begin{Vmatrix}
            x - \xi_{i+1}(t) \\
            w(t, x) - \omega_{i+1}(t)
        \end{Vmatrix} \leq
        \mathrm{e}^{(\hat T + \Delta_t) \hat F}
        \begin{Vmatrix}
            \Delta_x \\
            \Delta_w
        \end{Vmatrix}
    \end{equation*}
    where $w(t, x)$ is the exact solution of the problem (\ref{eq: general case}) and functions $\xi_i(t)$ and $\omega_i(t)$ are the outputs of the Algorithm.
\end{theorem*}

\begin{corollary*}
    Under the assumptions of the Theorem, the following holds for all $t\geq 0$ and $x \in [0, \ell]$:
    \begin{equation*}
        \big| w(t, x) - \tilde w(t, x) \big| \leq
        2 \,\mathrm{e}^{(\hat T + \Delta_t) \hat F}
        \begin{Vmatrix}
            \Delta_x \\
            \Delta_w
        \end{Vmatrix}
    \end{equation*}
    where $w(t, x)$ is the exact solution of the problem (\ref{eq: general case}) and $\tilde w(t, x)$ is the approximation of $w(t, x)$ obtained via linear or nearest-neighbor interpolation over the grid (\ref{eq: grid}).
\end{corollary*}

\begin{remark}
    The Theorem assumes that the PDE (\ref{eq: general case pde}) has the form (\ref{eq: Theo pde}) which excludes dependence of the dynamics on the full state of the system. Such full-state feedback may, in general, lead to divergence of the Algorithm and there would be no time-invariant accuracy estimation (although it can be established when $\hat T \hat F$ is small enough). This is due to an interpolant being used in place of the full PDE state in the characteristic ODEs (\ref{eq: characteristics algo}) which leads to unbounded error accumulation. Nonetheless, as can be seen in the example of Section \ref{sec: example}, the Algorithm may produce good results even in presence of full-state feedback.
\end{remark}

\begin{remark} \label{re: domain explanation}
    The Algorithm implies that the oldest ($N(t)$'th) characteristic lies outside the spatial domain $[0, \ell]$, i.e., $\xi_{N(t)} \geq \ell$ such that the interpolation grid (\ref{eq: grid}) covers the whole interval $[0, \ell]$. This is the reason for our assumption that the system (\ref{eq: general case}) is defined for all $x \geq 0$ instead of $x \in [0, \ell]$ (see Remark \ref{re: domain}).
\end{remark}

\begin{figure*}[t!]
    \begin{subfigure}{\textwidth}
        \centering
        \includegraphics[width=0.95\textwidth]{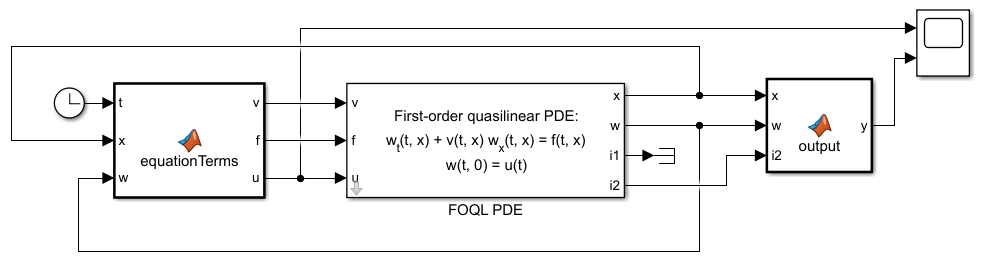}
        \subcaption{Simulink model of the dynamics (\ref{eq: general case}). The algorithm of Section \ref{subsec: simulation} is implemented in the \texttt{FOQL\;PDE} block.
        \label{fig: model}}
    \end{subfigure}
    \vspace{3mm}

    \begin{subfigure}{\textwidth}
        \centering
        \includegraphics[width=0.95\textwidth]{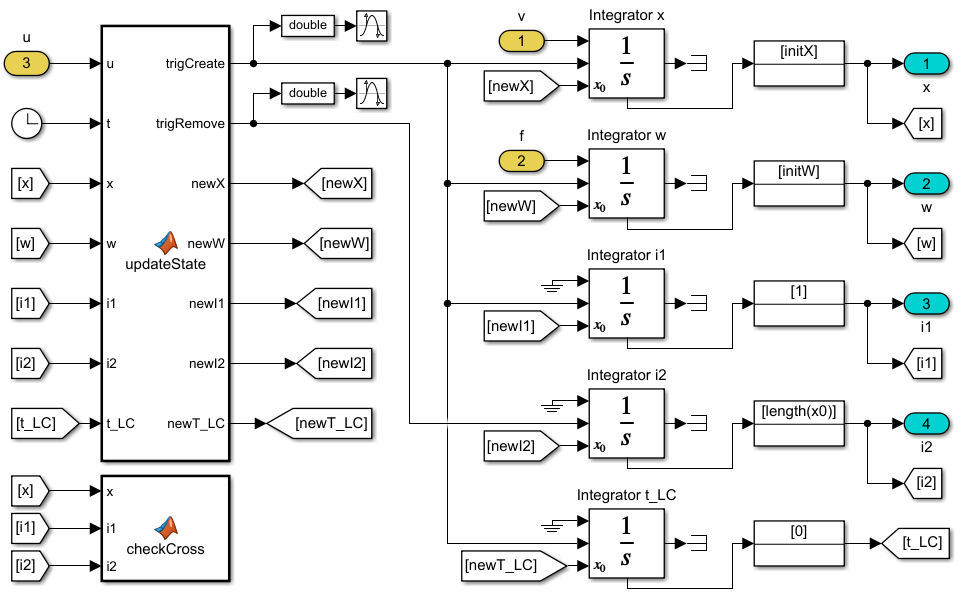}
        \subcaption{The inside structure of the \texttt{FOQL\;PDE} block.\label{fig: solver}} \vspace{3mm}
    \end{subfigure}
    \caption{Implementation of the method of characteristics in Simulink.}
\end{figure*}

\section{\uppercase{Implementation}}
\label{sec: implementation}

\noindent The above Algorithm has been realized in Simulink in the form of a masked subsystem \texttt{FOQL\;PDE} (``first-order quasilinear PDE'') which can be found in the library \texttt{PDECharactLib.slx} at \cite{githublink}. An example Simulink model (Fig.~\ref{fig: model}) which uses the subsystem can also be found there under the name \texttt{FOQL\_PDE\_SimpleExample.slx}. It implements (\ref{eq: general case}) with $\ell = 1$, velocity $v = x + 1.1 + \sin t$, source/sink term $f = -w(t,x)$, boundary input
\begin{equation*}
    u = \begin{cases}
        \frac{1}{2}(1 + \cos\frac{\pi t}{2}), & \text{ if } t\leq 16, \\
        0, & \text{ otherwise,}
    \end{cases}
\end{equation*}

\noindent
boundary output $y = w(t,1)$, and initial state function $w_0(x) = 1-x^2$. The reader is advised to download and inspect the model as description below is kept brief. The \textit{mask parameters} of the \texttt{FOQL\;PDE} block are:
\begin{itemize}
    \item ``Domain length (L)''~-- corresponds to $\ell$.
    \item ``Maximum number of characteristics (Nmax)''~-- hard upper bound on the number of characteristics $N(t)$ for the purpose of static memory allocation.
    \item ``x-values'' and ``w-values'' describe the initial function $w_0(x)$ by a grid of its values $(x, w_0(x))$.
    \item ``$\Delta$x'', ``$\Delta$w'', and ``$\Delta$t'' correspond to the parameters $\Delta_x$, $\Delta_w$, and $\Delta_t$ of the Algorithm.
    \item ``Tolerance for crossing characteristics''~-- sets the allowable upper bound on $\xi_i(t) - \xi_{i+1}(t)$. Crossing characteristics (i.e., the event $\xi_i(t) \geq \xi_{i+1}(t)$) indicate a shock wave in the solution which makes the method of characteristics not physically sound \cite{polyanin2002handbook}. Simulation will be terminated in this case with an error message. Setting the tolerance to a small positive number, firstly, helps to avoid termination caused by numerical inaccuracies rather than actual shock wave and, secondly, allows jumps in the solution. The jump is when $\xi_i(t) \equiv \xi_{i+1}(t)$ but $\omega_i(t) \neq \omega_{i+1}(t)$. At zero tolerance, this situation would be interpreted as two characteristics crossing though in fact it may be a valid discontinuous solution (see also Remark~\ref{re: initial jump}).
    \item The check box ``Terminate on overflow'', if checked, causes an error message when the current number of characteristics $N(t)$ equals its upper bound specified in the field ``Nmax'' above and the algorithm requires that another characteristic be created; otherwise, the simulation will silently continue without creating the characteristic. Unsetting the check box suppresses the characteristic creation trigger and may adversely affect the quality of the solution. The accuracy estimation given by the Theorem can no longer be guaranteed in this case.
\end{itemize}

The insides of the \texttt{FOQL\;PDE} block are exposed in Fig.~\ref{fig: solver}. The outputs \texttt{x} and \texttt{w} are constant-length vectors, their size specified in the mask parameter ``Nmax''. They contain the variable-length vectors $\Xi(t)$ and $\Omega(t)$ from (\ref{eq: Xi and Omega}) as sub-vectors. To facilitate adding and removing elements to and from $\Xi(t)$ and $\Omega(t)$ according to the Algorithm, vectors \texttt{x} and \texttt{w} are endowed with \textit{cyclic buffer} structure, meaning that the span of the elements containing $\Xi(t)$ and $\Omega(t)$ may wrap around the ends of \texttt{x} and \texttt{w}. The outputs \texttt{i1} and \texttt{i2} are the head and tail of the buffer, i.e., the indices at which the first characteristic $(\xi_1, \omega_1)$ and the last one $(\xi_{N(t)}, \omega_{N(t)})$ are stored in \texttt{x} and \texttt{w}.

Two topmost \textit{Integrator} blocks in Fig.~\ref{fig: solver} are used to solve the characteristic equations (\ref{eq: characteristics algo}), and the rest are employed to store \texttt{i1}, \texttt{i2}, and $t_{\textrm{LC}}(t)$. Creation and removal of the characteristics is done by triggering the \textit{Integrator} blocks' state reset.

\begin{figure}[t]
    \begin{subfigure}[b]{0.5\textwidth}
        \centering
        {\includegraphics[scale=0.5,trim=0cm 0.8cm 0cm 0cm,clip]{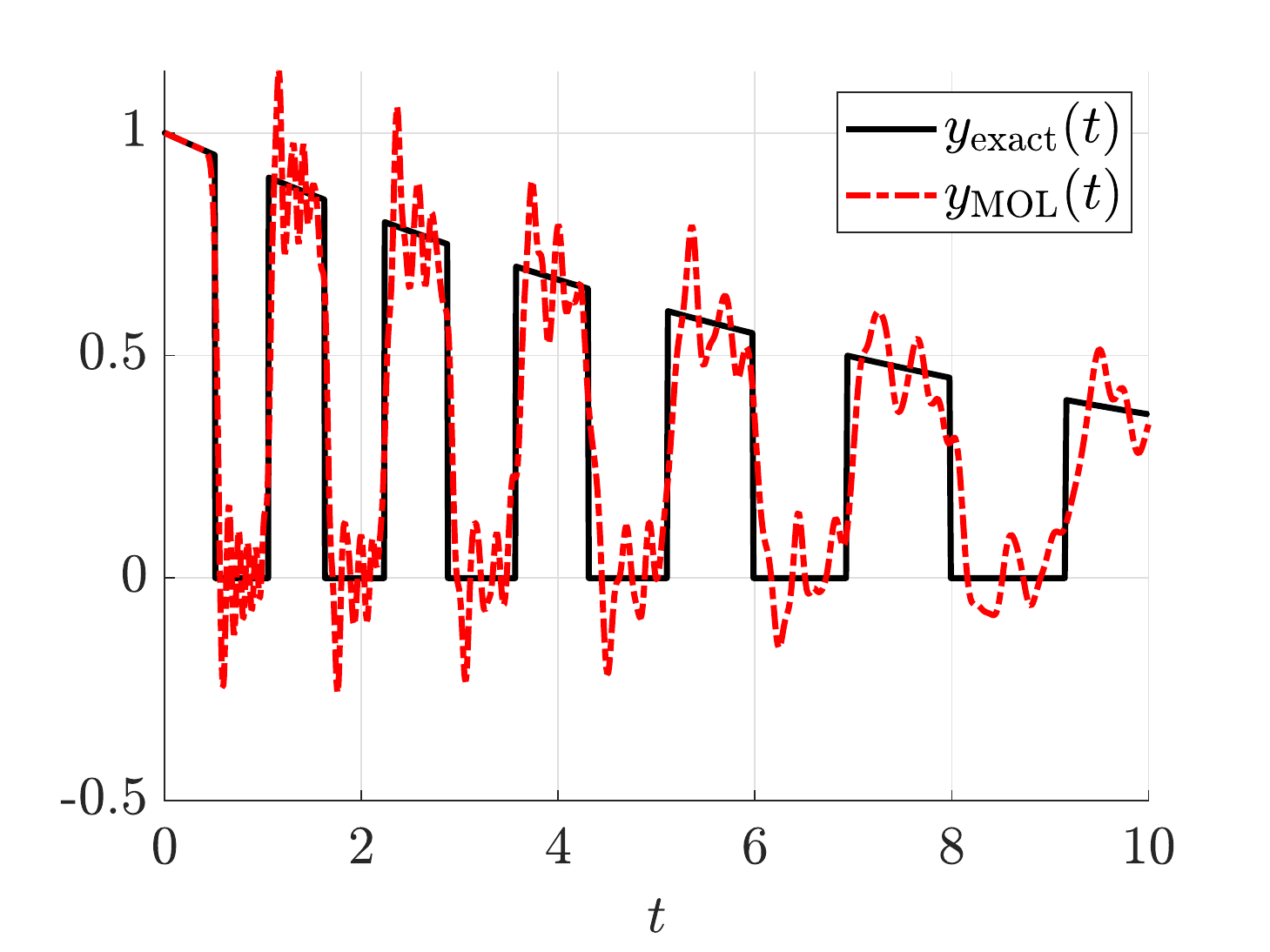}}
        \subcaption{$\gamma=-0.1$} \vspace{3mm}
        \label{fig: FDM_exact_case3}
    \end{subfigure}
    \begin{subfigure}[b]{0.5\textwidth}
        \centering
        {\includegraphics[scale=0.5,trim=0cm 0.8cm 0cm 0cm,clip]{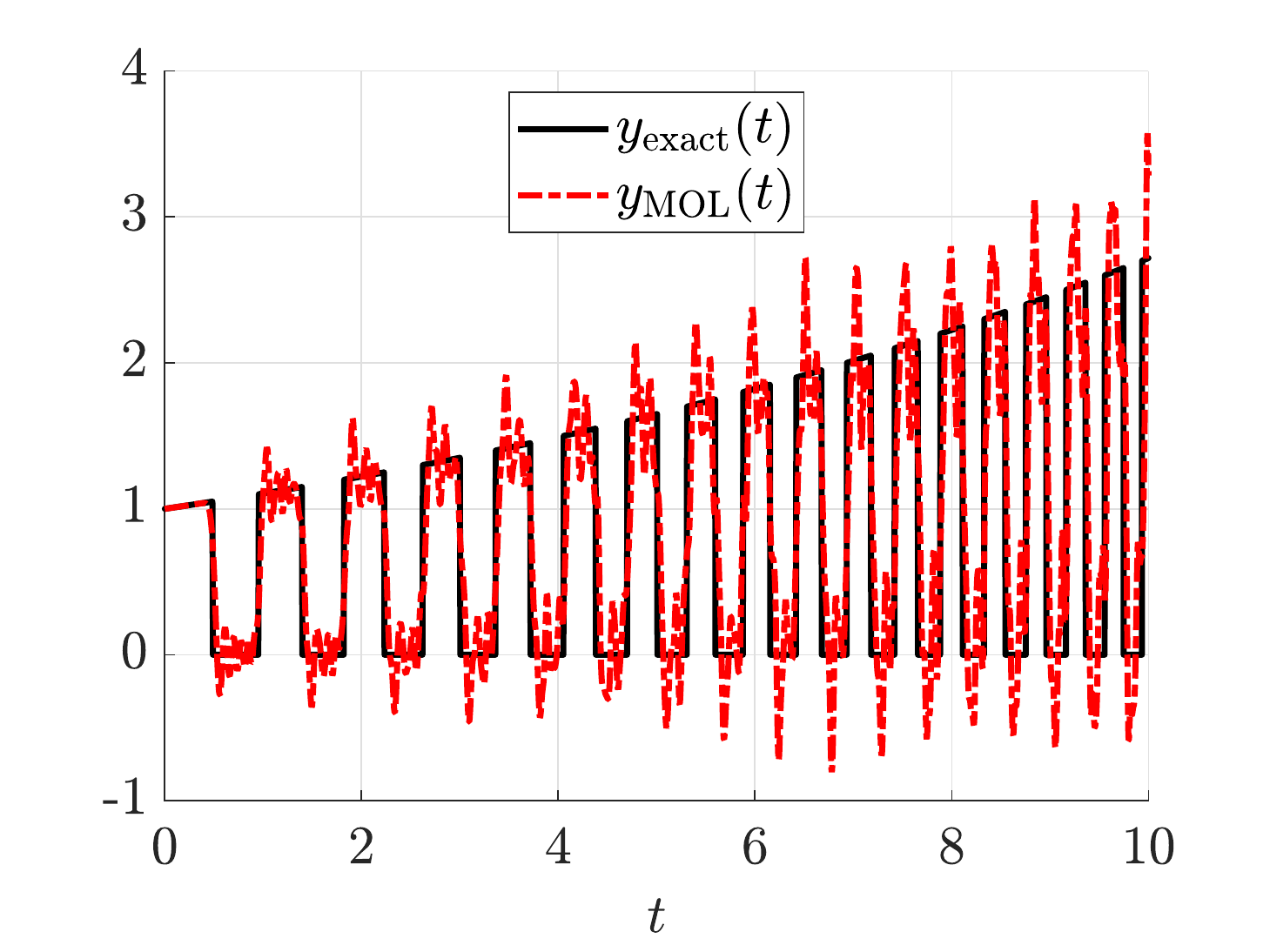}}
        \subcaption{$\gamma=0.1$} \vspace{3mm}
        \label{fig: FDM_exact_case3_unstable}
    \end{subfigure}
    \caption{Output $y(t)$ of (\ref{eq:Plant_StateFB})-(\ref{eq: example init}) obtained via the MOL with central finite differences and order $K=100$ compared to the exact solution (\ref{eq:case3}). The wiggles reveal numerical problems of the method.}
    \label{fig: exact_FDM}
\end{figure}

\begin{figure}[t!]
    \begin{subfigure}[b]{0.5\textwidth}
        \centering
        {\includegraphics[scale=0.5,trim=0cm 0.8cm 0cm 0cm,clip]{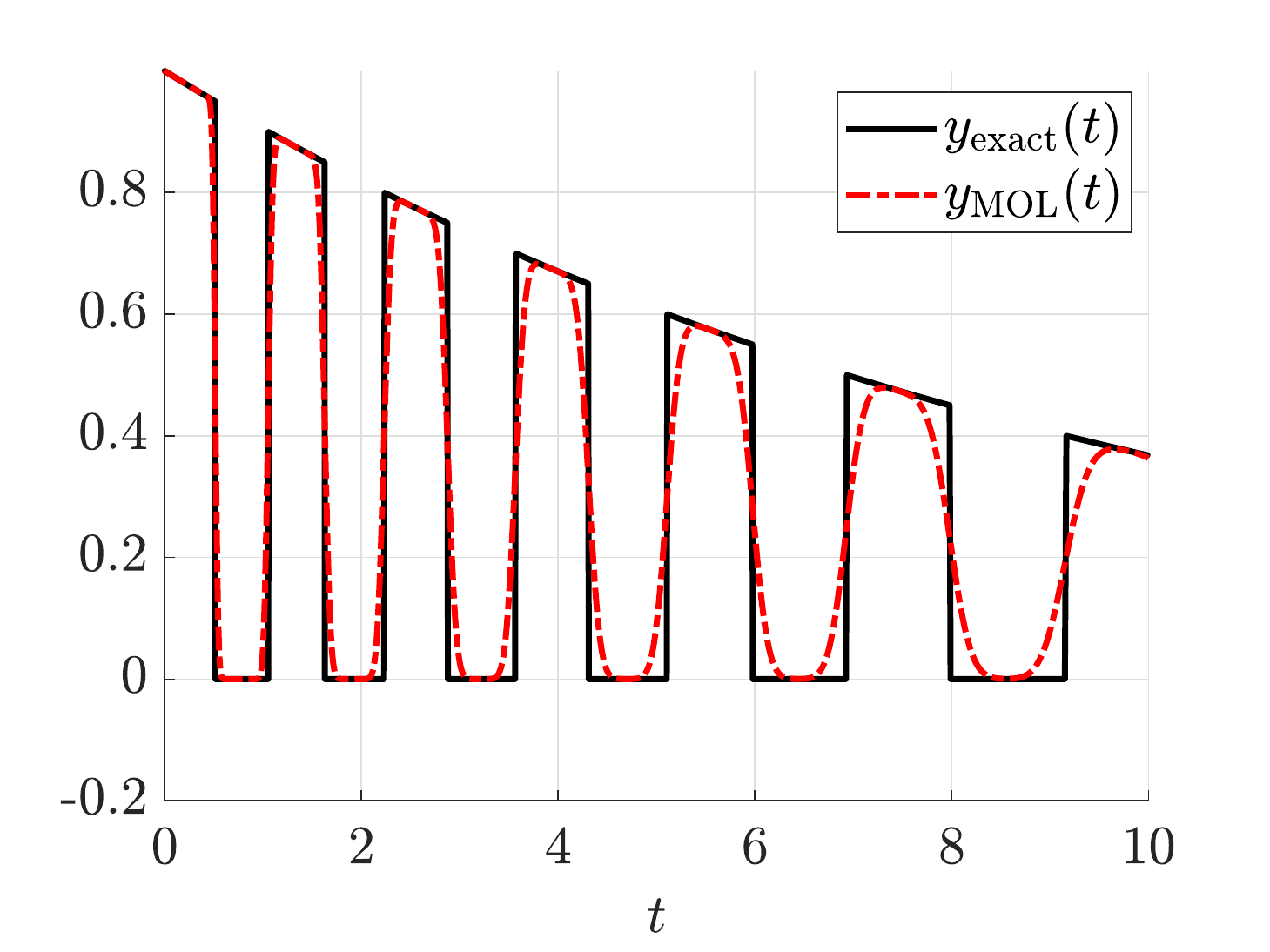}}
        \subcaption{$\gamma=-0.1$} \vspace{3mm}
        \label{fig: exact_FDM_case3_upwind_K1000}
    \end{subfigure}    
    \begin{subfigure}[b]{0.5\textwidth}
        \centering
        {\includegraphics[scale=0.5,trim=0cm 0.8cm 0cm 0cm,clip]{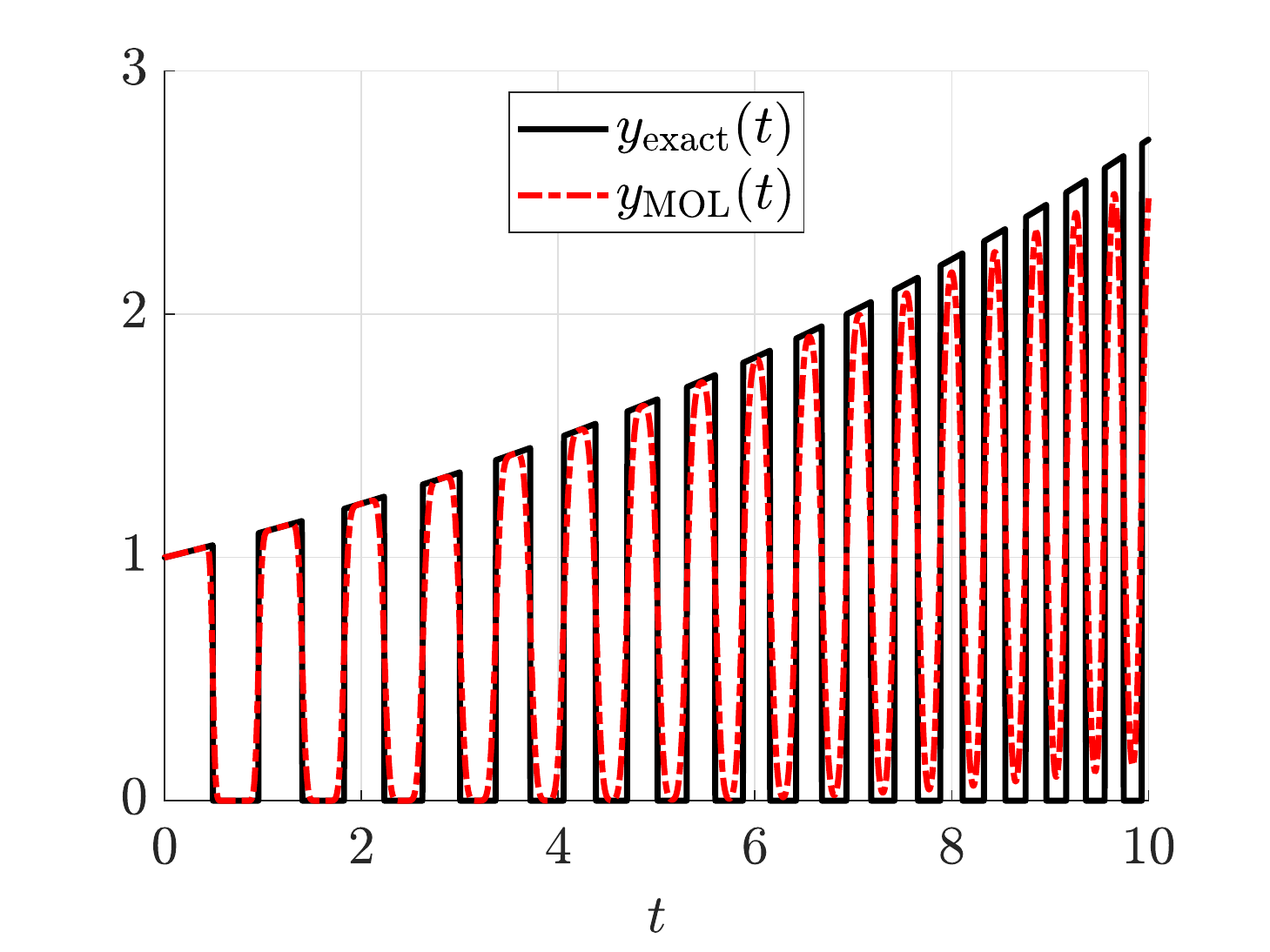}}
        \subcaption{$\gamma=0.1$} \vspace{3mm}
        \label{fig: exact_FDM_case3_unstable_upwind_K1000}
    \end{subfigure}    
    \caption{Output $y(t)$ of (\ref{eq:Plant_StateFB})-(\ref{eq: example init}) obtained via the MOL with upwind scheme and order $K=1000$ compared to the exact solution (\ref{eq:case3}). Numerical diffusion destroys the shape of the solution.}
    \label{fig: exact_FDM_upwind}
\end{figure}    

\begin{figure}[t!]
    \begin{subfigure}[b]{0.5\textwidth}
        \centering
        {\includegraphics[scale=0.5,trim=0cm 0.8cm 0cm 0cm,clip]{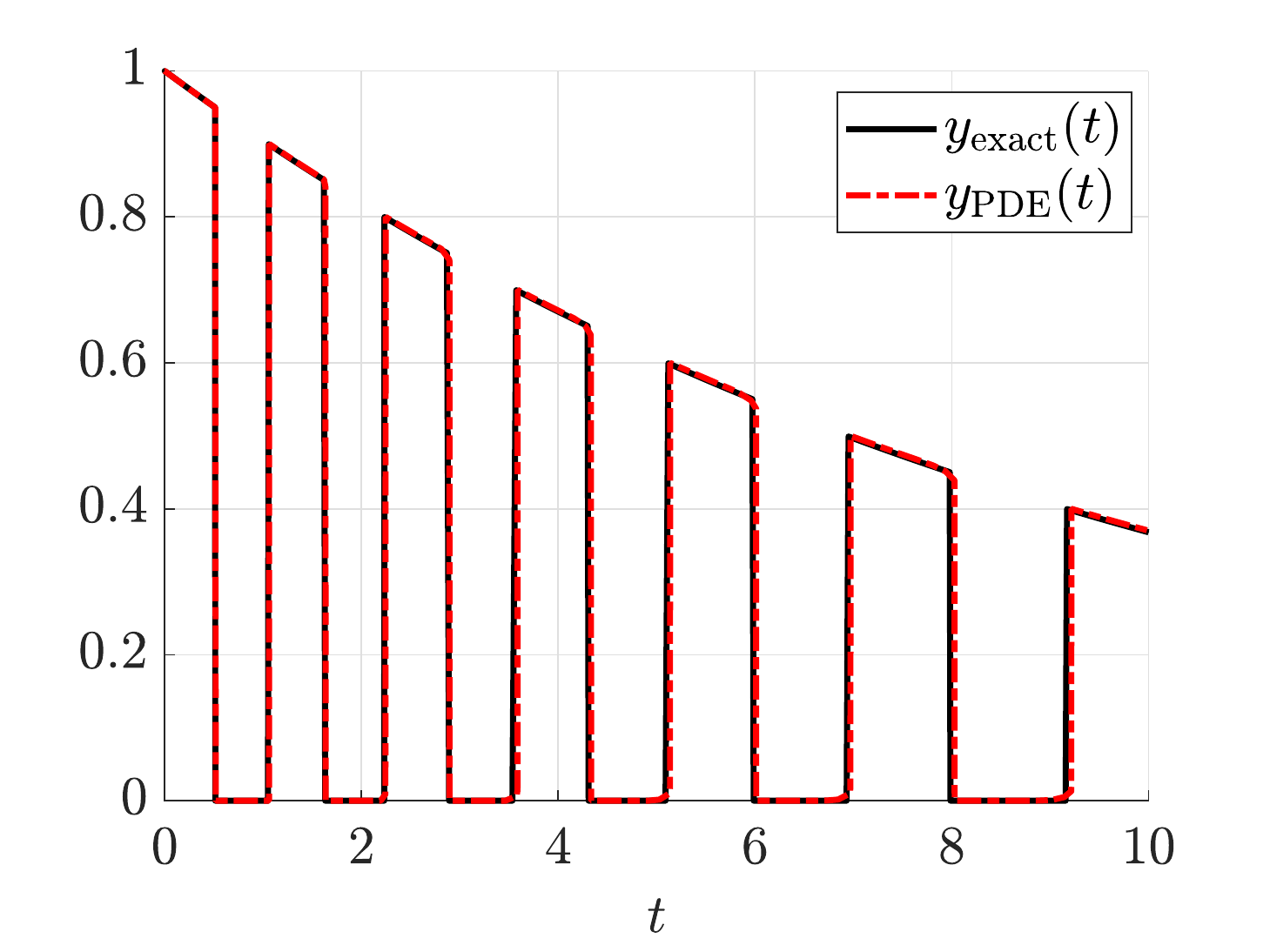}}
        \subcaption{$\gamma=-0.1$} \vspace{3mm}
        \label{fig: PDE_exact_case3}
    \end{subfigure}    
    \begin{subfigure}[b]{0.5\textwidth}
        \centering
        {\includegraphics[scale=0.5,trim=0cm 0.8cm 0cm 0cm,clip]{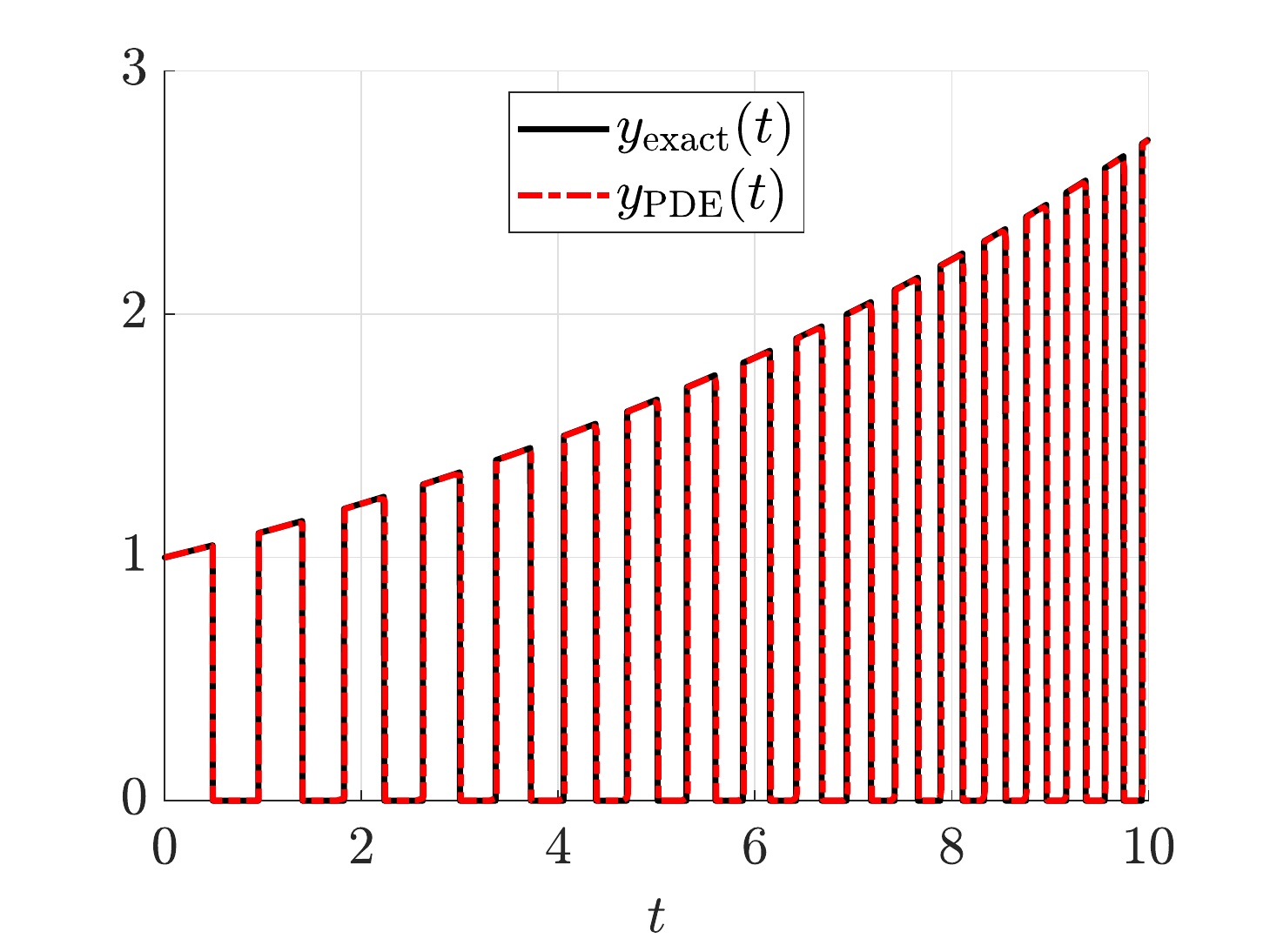}}
        \subcaption{$\gamma=0.1$} \vspace{3mm}
        \label{fig: PDE_exact_case3_unstable}
    \end{subfigure}        
    \caption{Output $y(t)$ of (\ref{eq:Plant_StateFB})-(\ref{eq: example init}) obtained via the proposed \texttt{FOQL\;PDE} block compared to the exact solution (\ref{eq:case3}). The disadvantages of the MOL are avoided to a large extent.}
    \label{fig: exact_PDE}
\end{figure}

\section{\uppercase{Example}}
\label{sec: example}

\noindent The repository \cite{githublink} contains three example models with the \texttt{FOQL\;PDE} block: \texttt{FOQL\_PDE\_SimpleExample.slx} is mentioned in the previous section, \texttt{FOQL\_PDE\_OutputFeedback.slx} is a real-life model of heat exchanger with output feedback control, and \texttt{FOQL\_PDE\_StateFeedback.slx} is an illustrative plant with full-state feedback. Here we discuss the latter and compare the solution produced by the \texttt{FOQL\;PDE} block to those obtained using the space discretization-based \emph{Method of Lines} (MOL).

The plant under consideration is
\begin{equation}
    \label{eq:Plant_StateFB}
    \begin{aligned}
        w_t + v\, w_x &= \gamma\; w(t,x),\quad x\in [0,1],\\
        w(0,x) &= w_0(x),\\
        w(t,0) &= u(t),\\
        \text{output: } y(t) &= w(t,1)
    \end{aligned}
\end{equation}
with a tunable constant $\gamma$ and full-state feedback
\begin{subequations}
    \label{eq:stateFB}
    \begin{align}
        v = v\big(w(t,\cdot)_{[0,1]}\big)
            &= 2\int\limits_0^1{w(t,x)\,\mathrm{d}x},
            \label{eq:stateFB_velocity}\\
        u(t) &= w(t,1). \label{eq:stateFB_input}
    \end{align}
\end{subequations}
Assuming the initial state is
\begin{equation} \label{eq: example init}
    w_0(x) = \begin{cases}
        0, & \text{if}\ x< \frac{1}{2}, \\
        1, & \text{otherwise,}
      \end{cases}
\end{equation}
\pagebreak

\noindent
one can obtain the exact output
\begin{equation} \label{eq:case3}
    y_{\mathrm{exact}}(t) = \begin{cases}
        a(t), & \text{if}\ s(t)- \lfloor s(t) \rfloor < 1/2,\\
        0, & \text{if}\ s(t)- \lfloor s(t) \rfloor\geq 1/2
    \end{cases}
\end{equation}
where $\lfloor \cdot \rfloor$ is the rounding-down operation and
\begin{equation*}
    a(t)=\mathrm{e}^{\gamma\, t}, \quad
    s(t)=\begin{cases}
        t, & \text{if}\ \gamma=0, \vspace{1mm} \\
        \dfrac{a(t)-1}{\gamma}, & \text{otherwise.}
        \end{cases}
\end{equation*}
For the purposes of simulation, the integral in (\ref{eq:stateFB_velocity}) is approximated using the trapezoidal rule.

As the first approach to simulating the dynamics (\ref{eq:Plant_StateFB})-(\ref{eq: example init}) we consider MOL with central finite difference approximations of the spatial derivative \cite{schiesser2012numerical} with $K$ discretization segments:

\begin{equation}
    \label{eq:MOL_Shang}
    \begin{gathered}
        w_{1,x}(t) \!:=\! w_x(t,x_1)
            \!\approx\! \frac{w_1(t) \!-\! 2w(t,0) \!+\! w_2(t)}
                {2{\scriptstyle \Delta}x}, \\
        w_{i,x}(t) := w_x(t,x_i)
            \approx \frac{w_{i+1}(t)-w_{i-1}(t)}
                {2{\scriptstyle \Delta}x}, \\
        w_{K+1,x}(t) := w_x(t,x_{K+1})
            \approx \frac{w_{K+1}(t)-w_K(t)}{{\scriptstyle \Delta}x} \\
        \big( i = 2,3,\ldots,K \big)
    \end{gathered}
\end{equation}
where $w_{i}(t):=w(t,x_i)$, $x_i := (i-1){\scriptstyle \Delta}x$ and ${\scriptstyle \Delta}x:=1/K$. Applying (\ref{eq:MOL_Shang}) to (\ref{eq:Plant_StateFB}) yields the ODE system
\begin{equation*} 
    \begin{aligned}
        \dot w_i(t) &= -v\, w_{i,x}(t) + \gamma\: w_i(t), \\
        w_i(0) &= w_0\left((i-1) {\scriptstyle \Delta}x \right) \\
        &\hspace{-1mm} \big( i = 1, 2, \ldots, K+1 \big).
    \end{aligned}
\end{equation*}
The results of this approach are shown in Fig.~\ref{fig: exact_FDM}.

The wiggles in the MOL solution appear because the initial function $w_0$ from (\ref{eq:case3}) is non-monotonic. Indeed, in general, finite difference-based MOL solutions suffer numerical diffusion, i.e., amplitude errors (smearing), and/or dispersion, i.e., phase errors (wiggles). In order to alleviate the effects, an upwind scheme for the spatial derivative, i.e.,
\begin{equation*}
    \begin{aligned}
        w_{1,x}(t) &:= w_x(t,x_1)
            \approx \frac{w_{1}(t)-w(t,0)}{{\scriptstyle \Delta}x}, \\
        w_{i,x}(t) &:= w_x(t,x_i)
            \approx \frac{w_{i}(t)-w_{i-1}(t)}{{\scriptstyle \Delta}x} \\
        &\phantom{:::}\big( i = 2,3,\ldots,K+1 \big)
    \end{aligned}
\end{equation*}
has been implemented \cite{Zijlema15}. The results are smooth (Fig.~\ref{fig: exact_FDM_upwind}), however, a high resolution of discretization is required ($K=1000$). Furthermore, the numerical solution departs from the exact one rather quickly due to numerical diffusion.

Finally, Fig.~\ref{fig: exact_PDE} presents the results of simulation using our characteristics-based \texttt{FOQL\;PDE} block. It yields better results than the MOL as neither numerical dispersion nor diffusion are apparent.

\begin{remark}
    Further improvement of the MOL solution with upwind scheme could be achieved by creating higher order, monotone, nonlinear upwind schemes using, e.g., the flux limiting technique \cite{Zijlema15}. However, flux limiters such as Minimod, Superbee, or Van Leerm limiter, are difficult to implement due to their nonlinearity.
\end{remark}

\begin{remark}
    Performance of the \texttt{FOQL\;PDE} block can be enhanced by adjusting the parameters of the block itself as well as the Simulink solver configuration. For instance, to obtain the results in Fig.~\ref{fig: exact_PDE} we had to set ``Max step size'' in Simulink's \emph{Solver Configuration Parameters} to 0.1 and reduce the \texttt{FOQL\;PDE} block's parameter ``$\Delta$w'' to 0.01.
\end{remark}    

\begin{remark} \label{re: initial jump}
    The discontinuous initial state (\ref{eq:case3}) has been defined by setting the \texttt{FOQL\;PDE} block parameter ``x-values'' to \texttt{[0, 0.5, 0.5, 1]} and ``w-values'' to \texttt{[0, 0, 1, 1]}. Thus, two characteristics are initialized at the same point of the initial boundary with different ``w-values'', specifically, $\xi_2(0) = \xi_3(0) = 0.5$, $\omega_2(0) = 0$, $\omega_3(0) = 1$. This approach lets one define initial state with a jump exactly. It is allowed thanks to non-zero ``Tolerance for crossing characteristics''. At zero tolerance simulation would fail immediately.
\end{remark}

To sum up, the finite difference-based MOL often yields inaccurate results due to the described numerical artifacts (diffusion and dispersion) whereas the results produced by the \texttt{FOQL\;PDE} block agree well with the exact solution.

\section{\uppercase{Conclusions}}
\label{sec: conclusion}

\noindent The method of characteristics has been implemented in Simulink in the form of a masked subsystem which can be set up to simulate a wide range of dynamics described by first-order quasilinear PDEs. The solution showed good accuracy surpassing that of the discretization-based method of lines.

We conclude that the subsystem is a viable option for simulation of convection-reaction dynamics in Simulink. In the future, the approach may be extended to a larger class of equations such as first-order nonlinear PDEs and higher-order hyperbolic PDEs.

\bibliographystyle{apalike}
{\small
\bibliography{biblio}}

\end{document}